\documentstyle[12pt]{article}

\def\ai{\'{\i}}
\def\itt{\int_{\tau_1} ^{\tau_2}}
\def\oq{\overline Q}
\def\pp{\pi_\phi}
\def\po{\pi_\Omega}
\def\om{\Omega}
\def\e3o{e^{3\Omega}}
\def\4o2b{e^{4\Omega+2\phi}}
\def\e2ob{e^{2(\Omega-\phi)}}
\def\la{\lambda}
\def\e-3{e^{-3\Omega}}

\def\e6ob{e^{6(\Omega+\phi)}}

\def\e3ob{e^{3(\Omega+\phi)}}
\def\be{\begin{equation}}
\def\ee{\end{equation}}

\def\px{\pi_x}
\def\ep{\epsilon}
\def\py{\pi_y}
\def\op{\overline P}
\begin{document}
\baselineskip.33in

\centerline{\large{\bf Global phase time and path integral}}

\centerline{\large{\bf for string cosmological models}}

\vskip0.8cm

\centerline{Gast\'on Giribet}

\centerline{\it  Instituto de Astronom\ai a y F\ai sica del Espacio}
\centerline{\it C.C. 67, Sucursal 28 - 1428 Buenos Aires, Argentina} 
\centerline{\it E-mail: gaston@iafe.uba.ar }

\bigskip

\centerline{Claudio Simeone}

\centerline{\it  Departamento de F\ai sica, Comisi\'on Nacional de
Energ\ai a At\'omica}
\centerline{\it Av. del Libertador 8250 - 1429 Buenos Aires, Argentina}
\centerline{\it and}
\centerline{\it Departamento de F\ai sica, Facultad de Ciencias Exactas y
Naturales}
 \centerline{\it  Universidad de Buenos Aires,  Ciudad Universitaria}
\centerline{\it Pabell\'on I - 1428, Buenos Aires, Argentina}
\centerline{\it E-mail: simeone@tandar.cnea.gov.ar}

\vskip1cm

\noindent A global phase time is identified for homogeneous and isotropic
cosmological models yielding from the low energy effective action of
closed bosonic string theory.    When the Hamiltonian constraint allows
for the existence of an intrinsic time,  the quantum transition amplitude
is obtained by means of the usual path integral procedure for gauge
systems.

\vskip1cm


\newpage

\noindent {\bf 1. Introduction}

\bigskip

\noindent An essential property of  gravitational dynamics  is that the
canonical   Hamiltonian  vanishes on the physical trajectories of the
system; the constraint ${\cal H}\approx 0$ reflects  that the evolution
is given in terms of a parameter $\tau$ which does not have physical
significance. This leads to a fundamental difference between ordinary
quantum mechanics and the  quantization of  gravitation, because the
existence of a unitary quantum theory is  related to the possibility of
defining the time as an absolute parameter. The identification of a global
phase time [1] can therefore be considered as the previous step before
quantization [2].

In the theory of gravitation the  Hamiltonian not only generates the
dynamical evolution, but it also acts as a generator of gauge
transformations which connect any pair of succesive points on each
classical trajectory of the system. While the dynamics is given by a
spacelike hypersurface evolving in spacetime, including arbitrary local
deformations which yield a multiplicity of times, the same motion can be
generated by gauge transformations [3]. It is therefore natural to think
that the gauge fixing procedure can be a way to identify a global time. 

However,  as the action of  gravitation is not gauge invariant at the
boundaries, this idea  could not be  used, in principle, to give a direct
procedure for deparametrizing mini\-superspaces: while ordinary gauge
systems admit canonical gauges $\chi(q^i,p_i,\tau)=0$,  only derivative
gauges would be admissible for cosmological models [4,5]. These gauges
cannot define a time in terms of the canonical variables. At the quantum
level this has the consequence that the usual  path integral  for ordinary
gauge systems could not be applied.

In the present paper we give a proposal for solving these problems in the
case of isotropic and homogeneous cosmological models resulting from the
bosonic closed string theory. We define a canonical transformation so that
the action of the   minisuperspaces is  turned into that of an ordinary
gauge system [6]; then we use  canonical gauge conditions to identify
a global phase time in terms of the canonical variables    for most
possible values of the parameters characterizing the models. When the
Hamiltonian has a potential with  a definite sign an intrinsic time $t(q)$
is defined, and the quantum transition amplitude for separable models is
obtained in the form of a path integral for an ordinary gauge system; the
$\tau-$dependent gauge choice used to identify the time determines the
time integration parameter and the  observables to be fixed at the end
points.  Differing from our previous analysis [6] (which was restricted
to simple models within the framework of general relativity),  now we also
obtain an extrinsic time $t(q,p)$ for the models, and, more important, we
give  a reduction procedure which leads to a conserved true Hamiltonian,
thus making more clear the meaning of the quantization.

\vskip0.8cm

\noindent {\bf 2. String cosmology models}

\bigskip

\noindent {\bf 2.1.}  {\it Gauge invariant action}

\medskip

\noindent The cosmological field equations yielding from the low energy
action of string theory show a remarkable T--duality symmetry  that
appears manifestly in terms of redefinition of the fields. The duality
properties of the models make string cosmology   very interesting, since
it makes possible to propose a pre--big bang phase for the universe [7].
The quantization of string cosmological models has been analyzed in the
context of the graceful exit problem (for a detailed discussion see
references [8] and [9], and references therein), and  it has been remarked
[10,11] that   a careful discussion  of the subtleties that are typical
of the quantization of gauge systems is required. In the present work  the
formal aspects of the problem are studied, and we give a solution for the
models whose  Hamilton--Jacobi equation is separable; we show that some
results are valid also for more general models. 

The massless states of bosonic closed string theory are the dilaton $\phi
$,
the two-form field $B_{\mu \nu }$ and the graviton $g_{\mu \nu }$ which
fixes the background geometry.
The low energy effective action that describes the long-wavelength limit
of
the massless fields dynamics is (written in Einstein frame)
\begin{equation}
S=\frac 1{16\pi G_D}\int d^Dx\sqrt{-g}\left( R-ce^{2\phi /(D-2)}-\frac
1{(D-2)}(\partial \phi )^2-\frac 1{12}e^{-4\phi /(D-2)}(dB)^2\right) 
\end{equation}
where $c= {2\over 3\alpha ^{\prime }}(D-26)$, being $\alpha ^{\prime }$
the
Regge slope, and $dB$ is the exterior derivative of the field $B_{\mu \nu
}$. In this paper we consider $c$ as an arbitrary real
parameter.

The Euler-Lagrange equations yielding from the action (1) admit
homogeneous
and isotropic solutions in four dimensions [12,13,14,15]. Such  solutions
present a Friedmann--Robertson--Walker form for the metric, namely 
\begin{equation}
ds^2=N^2(\tau )d\tau ^2-e^{2\Omega (\tau)}\left(
\frac{dr^2}{1-kr^2}+r^2d\theta
^2+r^2\sin ^2\theta d\varphi ^2\right).  \end{equation}
For the dilaton $\phi $ and the field strength ${\bf H}=dB$, the
homogeneity and
isotropy constraints demand 
\be
{\bf H}_{ijk} =\lambda\varepsilon _{ijk} \ \ \  \ \ \ \ \ \ \ \ \ 
\phi  = \phi (\tau),  \ee
where $\varepsilon _{ijk}$ is the volume form on the constant-time
surfaces
and $\lambda$ is a real number.
The Einstein frame action for this system in four dimensions is given by 
\be S={1\over 2}\int d^4x\sqrt{-g}Ne^{3\om}\left[-{{\dot\om}^2\over
N^2}+{{\dot\phi}^2\over
N^2}-2ce^\phi+\delta_{\lambda,0}ke^{-2\om}-\delta_{k,0}\lambda^2e^{-6\om-2\phi}\right],\ee
where the $\delta 's$ are introduced to consider the cases of a flat model
with two-form field different from zero, and a closed or open model with
${\bf H}_{\mu\nu\rho}=0$.
 If we put the action in the Hamiltonian form 
\be S=\int d\tau\left[\po\dot\om+\pp\dot\phi-N{\cal H}\right]\ee
 the canonical Hamiltonian is 
\be{\cal H}={1\over 2}e^{-3\om}\left(-\po^2+\pp^2+2c
e^{6\om+\phi}-\delta_{\lambda,0}ke^{4\om}+\delta_{k,0}\lambda^2e^{-2\phi}\right)\approx
0.\ee

As the  constraint  is quadratic in the momenta,  the action is not gauge
invariant at the boundaries of the trajectories [4,5,6]; however, if the
Hamilton-Jacobi  equation associated with ${\cal H}$ is separable, the
action  can be turned into that of an ordinary gauge system  by improving
it with gauge invariance at the end points [6]. 
 We shall begin by considering the following generic form for the scaled
Hamiltonian  $H\equiv 2e^{3\om}{\cal H}:$
\be H= -\po^2+\pp^2+4Ae^{n\om+m\phi}\approx 0,\ee
where $A$ is an arbitrary real constant and $m\neq n$. In general, this
Hamiltonian is not separable in terms of the original canonical variables.
Then we define
\be x\equiv\left({2\over n+m }\right) e^{(n+m)(\om+\phi)/2}, \ \ \ \ \ \
y\equiv \left({2\over n-m}\right) e^{(n-m)
(\om-\phi)/2}\ee
so that dividing $H$ by $(n^2-m^2)xy>0$ we can define the equivalent
constraint
\be H'\equiv {H\over (n^2-m^2)xy}=-\px\py+A\approx 0.\ee
The solution of the corresponding  Hamilton--Jacobi equation $-(\partial
W/\partial x)(\partial W/\partial y) + A =E'$  which is obtained by
matching 
 the integration constants $\alpha,E'$ to the momenta $\op,\op_0$ is
$$W(x,y,\op_0,\op)=\op x+y\left({ A- \op_0\over \op}\right).$$
 The solution $W$  generates a canonical transformation $(q^i,p_i) \to
(\oq^0,\oq,\op_0,\op)$ 
which identifies $ H'$ with  $\op_0$. The variables $(\oq,\op)$  are
conserved observables because $[\oq, H']=[\op, H']=0$, so that they would
not be appropriate to characterize the dynamical evolution.
The function
 \be F=P_0\oq^0+f(\oq ,P,\tau) \ee
 generates a second transformation in the space of observables $(\oq,\op)
\to (Q,P)$,
such that the new Hamiltonian $K=NP_0+\partial f/\partial\tau$ does not
vanish, and $Q$ is a non conserved observable because $[Q,H']=[P, H']=0$
but $[Q, K]\neq 0$. For $Q^0$ we have
$ [Q^0,H']=[Q^0,P_0]=1,$
and then $Q^0$ can be used to fix the gauge [16]. If we choose
\be
f=\oq P+T(\tau)/ P\ee
with $T(\tau)$ a monotonic function then the new canonical variables are
given by
$$P_0=-\px\py+A,\ \ \ \ \ \ \  \ P=\px,$$
\be
Q^0=-{y\over P},\ \ \ \ \ \ \ \ Q=x-\left({y(A- P_0)+ T(\tau)\over
P^2}\right) \ee
($P=\px$ cannot be zero on the constraint surface).
 The coordinates and momenta $(Q^i,P_i)$ describe an  ordinary gauge
system with a constraint $P_0=0$ and a true Hamiltonian $\partial
f/\partial \tau=(1/P)(dT/d\tau)$ which conmutes with $K$. Its action is
\be {\cal S}[Q^i,P_i,N]=\itt \left( P{dQ\over d\tau} +P_0{dQ^0\over
d\tau}-NP_0 - {1\over P} {dT\over d\tau}\right)d\tau .\ee
If we write ${\cal S}$ in terms of the original variables we must add end
point terms [17,6] of the form [6] $ B= \oq^i\op_i -W+Q P-f$ so that 
\be {\cal S}[\om,\phi,\po,\pp,N]=\itt \left(\pp{d\phi\over
d\tau}+\po{d\om\over d\tau}-N{\cal H}\right)d\tau + B(\tau_2) - B(\tau_1),
\ee
where
$$ B(\tau)  =  {1\over (\po+\pp)}\left[{\pp^2-\po^2+4Ae^{n\om+m\phi}\over
n-m}   + 4Ae^{(n+m)(\om+\phi)/2}
\left({2e^{(n-m)(\om-\phi)/2}\over n-m}+ {T(\tau)\over A}\right)\right].$$
As $\px=\op=(1/2)(\po+\pp)e^{-(n+m)(\om+\phi)/2}$ we can write
$$B(\tau)= -Q^0 P_0 -2A\left(Q^0-{T(\tau)\over A\op}\right).$$
Under a gauge transformation generated by $\cal H$ we have $\delta_\ep
B=-\delta_\ep S$, so that the action $\cal S$ is effectively endowed with
gauge invariance over the whole trajectory and canonical gauge conditions
are admissible.

\bigskip

\noindent {\bf 2.2.} {\it Extrinsic time}

\medskip

A global phase time $t$ must verify $[t,{\cal H}]>0$ [1], but as   ${\cal
H}={\cal F}(\om,\phi) H'={\cal F}(\om,\phi)P_0$ with ${\cal F}>0$, then if
$t$ is a global phase time we also have $[t,P_0]>0$. Because
$[Q^0,P_0]=1$, an extrinsic time can be identified by imposing a
$\tau-$dependent gauge  of the form
$ \chi\equiv Q^0-T(\tau)=0$ and defining $t\equiv T.$ We  then obtain
\be t(\om,\phi,\po,\pp)  =  Q^0
 =  {4e^{n\om +m\phi}\over (m-n)(\po+\pp)}.\ee
  Using the constraint equation (7) we can write $
t(\po,\pp)=(n-m)^{-1}(\pp-\po) /A.$
For the scaled constraint $ H=2e^{3\om}{\cal H}$ with  $k=\lambda=0$  we
have
$4A=2c,\ n=6,\ m=1$. Then the extrinsic time is 
\be t(\om,\phi,\po,\pp)  =  -{4e^{6\om+\phi}\over 5(\po+\pp)}.\ee 
We can go back to the   constraint $ H$ with $k\neq 0$ and evaluate
$[t,H]$.  For an open model ($k=-1$) a simple calculation gives that
$[t,H]>0$ for both $c<0$ and $c>0$. For the case $k=1$, instead, an
extrinsic global  phase time is $t(\po,\pp)=(2/5c)(\pp-\po)$ if $c<0$.   

 In the case of  the scaled  constraint  with $c=k=0$ we have
$4A=\lambda^2,\ n=0,\ m=-2,$ and the extrinsic time reads
\be t(\om,\phi,\po,\pp)  =  -{2e^{-2\phi}\over \po+\pp}.\ee 
If we then consider $c\neq 0$ and we compute the bracket $[t,H]$ we find
that this is positive definite if $c<0.$ Hence  the time given by (17) is
a global phase time for this case. 

\bigskip

\noindent {\bf 2.3.} {\it Intrinsic time and path integral}

\medskip

\noindent The action (13) can be used to compute the amplitude for the
transition $|Q_1,\tau_1>\, \to\, |Q_2,\tau_2>$ ($Q^0$ is a spurious degree
of freedom for the gauge system) 
 by means of a path integral in the form
\be <Q_2,\tau_2|Q_1,\tau_1>=\int DQ^i DP_i  DN \delta(\chi )\,\vert
[\chi,P_0]\vert\, \exp\left[ i \itt \left( P_i{dQ^i\over
d\tau}-NP_0-{\partial f\over\partial\tau}\right) d\tau\right].\ee
Here $\vert [\chi, P_0]\vert$ is the Fadeev-Popov determinant; because the
constraint is simply $P_0=0$, canonical gauges are admissible. But what we
want to obtain is the amplitude $<q_2^i|q_1^i>$, so that we  should show
that both amplitudes  are equivalent. This is fulfilled if the paths are
weighted in the same way by ${\cal S}$ and $S$ and if $Q$ and $\tau$
define a point in the original configuration space, that is, if  a state
$|Q,\tau>$ is equivalent to $|q^i>$. This is true only if there exists a
gauge such that $\tau=\tau(q^i)$, and such that on the constraint surface
the boundary terms in (14) vanish [6]. 

The existence of  a gauge condition yielding  $\tau=\tau(q^i)$ is closely
related to the existence of an intrinsic time [18]. A (globally good)
gauge such that $\tau=\tau (q^i)$ should be given by a function $\chi
(q^i,\tau)=0$ fulfilling $[\chi,{\cal H}]\neq 0$, while a function
$t(q^i,p_i)$ is  a global phase time if
$ [t,{\cal H}]>0.$
Because the supermetric $G^{ik}$ does not depend on the momenta, a
function $t(q)$ is a global phase time if the bracket
$$[t(q),{\cal H}]  = [t(q), G^{ik}p_ip_k]
 =  2{\partial t\over\partial q^i} G^{ik}p_k$$
is positive definite. For a constraint whose potential can be zero for
finite values of the coordinates, the momenta $p_k$ can be all equal to
zero at a given point, and $[t(q),{\cal H}]$ can vanish. Hence an
intrinsic time can be identified only if the potential  has a definite
sign. 

On the constraint surface $H'=P_0=0$ the terms $B(\tau)$ clearly vanish in
the canonical gauge
\be \chi\equiv  Q^0 - {T(\tau)\over A\op}=0\ee
which is equivalent to
$T(\tau)=2(m-n)^{-1}A e^{(n-m)(\om-\phi)/2},$   and then it defines
$\tau=\tau(\om,\phi).$  As $P=\op$ and thus $Q^0\op= -y(\om,\phi)$, an
intrinsic time $t$ can be defined   as  
$$t\equiv{\eta T\over 2A}$$ 
if we apropriately choose  $\eta.$ 
 We have 
$[t,H']=(\eta / 2)[Q^0 \op,P_0]=(\eta / 2) \op,$
and because $\op=\px$ then to ensure that $t$ is a global phase time we
must choose $\eta= sign(\px )=sign(\po+\pp).$ 

In the case $A>0$ it is $|\po|>|\pp|$ (so that $sign(\px)=sign(\po)$)  and
the constraint surface splits into two disjoint sheets identified by the
sign of $\po$; in the case $A<0$ it is $|\pp|>|\po|$  and the two sheets
of the constraint surface are given by the sign of $\pp.$ Hence in both
cases $\eta$ is determined by the sheet of the constraint surface  on
which the system evolves; we therefore have that for $A>0$ the intrinsic
time can be written as 
\be t(\om,\phi)= \left({1\over m-n}\right) sign(\po)
e^{(n-m)(\om-\phi)/2},\ee
while for $A<0$ we have 
\be t(\om,\phi)= \left({1\over m-n}\right) sign(\pp)
e^{(n-m)(\om-\phi)/2}.\ee
For the constraint with $k=\lambda=0$ the intrinsic time is 
$$ t(\om,\phi)= -{1\over 5} sign(\po) e^{5(\om-\phi)/2}\ \ \ \ \ \
\mbox{if}\ \ \ \ \ \ c>0,$$
and 
$$ t(\om,\phi)= -{1\over 5} sign(\pp) e^{5(\om-\phi)/2}\ \ \ \ \ \
\mbox{if}\ \ \ \ \ \ c<0.$$
By evaluating the bracket $[t,H]$ for $H$ with $k\neq 0$ we find that the
intrinsic time obtained in the case $c>0$ is also a time for an open model
($k=-1$), and the time for $c<0$ is a time also for $k=1.$

In the case of the constraint  with $c=k=0$ we obtain 
$$ t(\om,\phi)= -{1\over 2} sign(\po) e^{(\om-\phi)},$$
and a simple calculation shows that this is also a global phase time for a
more general model with $c>0.$
 
 Because we have shown that there is a gauge such that $\tau=\tau (q^i)$
and which makes the endpoint terms vanish, we can obtain the amplitude for
the transition $|\om_1,\phi_1>\, \to\, |\om_2,\phi_2>$ by means of a path
integral in the variables $(Q^i,P_i)$ with the action (13). This integral
is gauge invariant, so that we can compute it in any canonical gauge.
According to (18), on the constraint surface $P_0=0$ and with the gauge
choice (19),   the transition amplitude is 
\be <\phi_2,\om_2\vert\phi_1,\om_1 >=\int DQDP\exp\left[i\int_{T_1} ^{T_2}
\left(PdQ-{1\over P}dT\right) \right],\ee
where the end points are given by $$T_a=\left({2A\over  m-n}\right)
e^{(n-m)(\om_a-\phi_a)/2}$$
($a=1,2$); because on the constraint surface and in gauge (19) the true
degree of freedom reduces to $Q=x$, then the boundaries of the paths in
phase space are  
$$Q_a=\left({2\over n+m}\right) e^{(n+m)(\om_a+\phi_a)/2}.$$
For the Hamiltonian   with $\lambda=0$ and null curvature the end  points
are given by
$T_a= -(c/5) e^{5(\om_a-\phi_a)/2}$, while $Q_a=(2/ 7)
e^{7(\om_a+\phi_a)/2}.$
In the case of  $c=k=0$ we have $T_a= -(\lambda^2/ 4) e^{(\om_a-\phi_a)}$
and
$Q_a=-e^{-(\om_a+\phi_a)}.$

After the gauge fixation we have obtained  the path integral  for a system
with one physical degree of freedom.  A point to be remarked is that in
our previous work [6] the reduction procedure yielded  a true Hamiltonian
which was analogous to that of a massless particle with a time dependent
potential;  such a potential leads to particle creation, so that the
meaning of the minisuperspace quantization would not be completely clear.
Here, instead, we have avoided this difficulty because we have obtained a
true Hamiltonian $1/P$ which does not depend on time (see eq. (22)).

\vskip0.8cm

\noindent{\bf 3. Discussion}

\bigskip

 \noindent We have been able to use canonical gauge conditions for
deparametrizing homogeneous and isotropic cosmological models coming from
the low energy dynamics of bosonic closed string theory and,
simultaneously, to obtain the quantum transition amplitude in a  simple
form which clearly shows the separation between true degrees of freedom
and time. 

 We have analized models of two types: 1) models with homogeneous dilaton
field and vanishing antisymmetric $B_{\mu\nu}$ field ($\la=0$); 2) models
representing flat universes ($k=0$) with homogeneous dilaton and non
vanishing antisymmetric field. For  the cases considered we have been able
to identify a global phase time.
In the cases $\la=0,\ k=0,\ c\neq 0$ and $\la\neq 0,\ k=0,\ c=0$ the
Hamiltonian is easily separable and the potential has a definite sign.
Thus, an intrinsic time can be found and the quantum transition amplitude
is obtained by means of a path integral in  the new variables $(Q^i,P_i)$
describing an ordinary gauge system. The canonical gauge used to define
the time determines the integration parameter and the variables to be
fixed at the boundaries. 

Once we have found a time $t$ for the inmediately separable models, we
have identified the extended region of the parameter space where $t$ is a
global phase time. In fact, a  simple prescription can be given to
determine whether an extrinsic time for a system described  by a given
Hamiltonian  is also a time for a system described by a more general
constraint.  We have defined $ H'=g^{-1}(q)H$ with $g>0$, and because we
matched $P_0\equiv H'$, then  $t\equiv Q^0$ fulfills $[t,H']=1$ (and then
$ [t,H]=g>0$ on the surface $H=0$).
 If we consider an extended constraint $\tilde H=g(q)H'+ h$ and we
calculate the bracket of $t$ with $\tilde H$ we obtain 
$$[t,\tilde H]= g+H'[t,g]+[t,h].$$ 
Using that $\tilde H\approx 0$ we have that the condition
$$[t,\tilde H]=g-g^{-1}h[t,g]+[t,h]>0$$
must hold on the (new) constraint surface if $t$ is a time for the system
described by $\tilde H.$  For the system associated to the constraint (7),
from (8) and (9) we have that $g=4e^{n\om+m\phi}$;  if we add a term of
the form $h=\alpha e^{r\om+s\phi}$ to $H$ the condition turns to be
$$\alpha {e^{r\om+s\phi}\over (\pp+\po)^2}\left[{(n+m)-(r+s)\over
n-m}\right]> -1.$$

We have  restricted our analysis to the formal aspects of minisuperspace
quantization. A complete discussion about the limits of such approximation
as well as an analysis of the application of our method to the graceful
exit problem would require a detailed knowledge of the effective potential
for the dilaton. If the effective potential leads to a separable
Hamilton--Jacobi  equation the application of our procedure would result
straightforward. The minisuperspaces that we have quantized admit an
intrinsic time; however, an intrinsic time can be defined  only if the
constraint surface splits into two disjoint sheets. If the complete
Hamiltonian including the effective potential is separable but admits only
an extrinsic time, the variables to be fixed at the boundaries in the path
integral should involve not only the original coordinates but also the
momenta; this point would require a further discussion.

\vskip0.8cm

\noindent{\bf Acknowledgment}

\bigskip

This work was supported by CONICET and CNEA (Argentina).

\newpage

\noindent{\bf References}

\bigskip

\noindent 1. P. H\'aj\ai cek, Phys. Rev.  {\bf D34}, 1040 (1986).

\noindent 2. R. Ferraro, Gravit. Cosmol. {\bf 5},  195 (1999).

\noindent 3. A. O. Barvinsky, Phys. Rep. {\bf 230}, 237 (1993).

\noindent 4. C. Teitelboim, Phys. Rev. {\bf D25}, 3159 (1982).
 
\noindent 5. J. J. Halliwell, Phys. Rev.  {\bf D38}, 2468 (1988).

\noindent 6. H. De Cicco  and C. Simeone,  Int. J. Mod. Phys. {\bf A 14},
5105 (1999).

\noindent 7. G. Veneziano, Phys. Lett. {\bf B265}, 387 (1991).

\noindent 8.  M. Gasperini,  Class. Quant. Grav. {\bf 17} R1 (2000). M.
Gasperini, in {\it Proceedings of the 2nd SIGRAV School on Gravitational
Waves in Astrophysics, Cosmology and String Theory, Villa Olmo, Como},
edited by V. Gorini, hep-th/9907067.

\noindent 9. G. Veneziano, {\it String Cosmology: The pre-big bang
scenario}, Lectures delivered in Les Houches (1999), hep-th/0002094. 

\noindent 10. M. Cavagli\`a and A. de Alfaro, Gen. Rel. Grav. {\bf 29},
773 (1997).

\noindent 11. M. Cavagli\`a and C. Ungarelli, Class. Quant. Grav. {\bf
16}, 1401 (1999).

\noindent 12.  I. Antoniadis, C. Bachas, J. Ellis and D. V.
Nanopoulos, Phys. Lett.  {\bf B221}, 393, (1988). 

\noindent 13. A. A. Tseytlin, Class. Quant. Grav.
{\bf 9}, 979, (1992). 

\noindent 14. A. A. Tseytlin and C. Vafa, Nucl. Phys.  {\bf B372}, 443,
(1992). 

\noindent 15. D. S. Goldwirth and M. J. Perry, Phys. Rev.  {\bf D49}, 5019
(1994).

\noindent 16.  M. Henneaux  and C. Teitelboim, {\it Quantization of Gauge
Systems} (Princeton University Press, New Jersey, 1992).

\noindent 17.  M. Henneaux, C. Teitelboim  and J. D. Vergara, Nucl. Phys.
{\bf B387}, 391 (1992).

\noindent 18.  K. V. Kucha\v r, in {\it Proceedings of the 4th Canadian
Conference on General Relativity and Relativistic Astrophysics}, edited by
G. Kunstatter, D. Vincent and J. Williams (World Scientific, Singapore,
1992).

\end{document}